\documentclass[a4paper]{jpconf}
\usepackage{graphicx}
\usepackage{wrapfig}
\usepackage{subfigure}

\begin{document}
\title{Azimuthal Anisotropy in U+U Collisions at $\sqrt{s_{NN}} = 193 $ GeV with STAR Detector  at RHIC}

\author{ Yadav Pandit (for the STAR Collaboration)}

\address{Department of Physics, UIC, Chicago, USA}

\ead{ypandit@uic.edu}

\begin{abstract}
We report the measurement of  first (\textit {n} = 1) and higher order(\textit{n} = 2-5) harmonic coefficients  ($v_{n}$)  of the azimuthal anisotropy  in the distribution of the particles produced  in  
U+U collisions at  $\sqrt{s_{NN}}$ =193 GeV, recorded  with the STAR detector at RHIC.  The differential measurement of $(v_{n})$  is  presented
 as a function of transverse momentum ($p_{T}$) and centrality. We also present  $v_{n}$  measurement  in the ultra-central collisions. 
These data may provide strong constraints on the theoretical models of the initial condition in heavy ion collisions and the transport properties of
the produced medium. 
\end{abstract}


\section{Introduction}
      
  The study of azimuthal anisotropy, based on Fourier coefficients, is widely recognized as an important tool to probe the hot, dense matter created in heavy ion collisions ~\cite{methodPaper}.  These measurements  are sensitive  to early stage evolution  of  the system. In a picture of hydrodynamic expansion of the system formed in the collisions, these final state momentum anisotropies are expected to rise due to the initial state anisotropy driven  pressure gradients and subsequent interactions of the constituents. Specially the differential measurement of   azimuthal  anisotropy  have been found to be sensitive to initial condition, thermalization, the equation of state and transport coefficients of the medium  in the heavy ion collisions. Azimuthal  anisotropy  quantified by the Fourier coefficient of the particle distribution can be written as

\begin{equation}
E  \frac {d^{3} N}  {d^{3} p} = \frac {1 } { 2 \pi }  \frac{ d^{2} N } {p_{T} d p_{T} dy}  (1+\sum_{n=1}^{\infty}  {2v_{n} \cos n(\phi-\psi_{R})} .
\label{eq1}
\end{equation} 
 where  $p_{T}$ is the transverse momentum,  $y$ is rapidity and  $\phi $ is the azimuthal angle of each particle  and  $ \psi_{R}$ is the true reaction plane angle, defined by beam direction and impact parameter vector between two colliding nuclei. The sine terms in the Fourier expansion vanish due to the reflection symmetry with respect to the reaction plane. The reaction plane $ \psi_{R}$  is not
measurable directly a priori, so the Fourier  coefficients  are determined with respect to the estimated participant event planes~\cite{methodPaper},    

 \begin{equation}
v_{n}=\langle\cos n(\phi- \Psi_{n}) \rangle.
\end{equation}

where  $\Psi_{n}$ are the generalized participant event planes at all orders for each event. The estimated event plane from detected particles is  corrected for event plane resolution. The measured $v_{n}$ with respect to $\Psi_{n}$ for reasonable event-plane resolutions  are closer to the root-mean-square values~\cite{eventPlane} than the mean values.  The  first harmonic coefficient $v_{1}$, called directed flow, has two components rapidity odd directed flow generated due to bounce off motion of  the nuclei and rapidity even directed flow due to dipole asymmetry in the initial geometry. The directed flow measurement  should suppress the the contribution from global momentum conservation especially at peripheral collisions along with other non-flow contributions.  The rapidity odd $v_{1}(y)$  component at 200 GeV Au+Au collision is  found  very small at mid-rapidity and becomes significant at  forward rapidity ~\cite{WW2010Proceeings}. The rapidity even component  $v_{1}(y)$ recently reported at 200 GeV Au+Au collisions  ~\cite{DipoleHQProceedings} form STAR collaboration is found  to have weak rapidity dependence as predicted by some models~\cite{Luzum}. The transverse momentum dependence of rapidity event $v_{1}(p_{T})$ is of interest in the present study.   The second harmonic coefficient $v_{2}$, popularly known as elliptic flow have been extensively studied both experimentally and theoretically. Recently higher order harmonics have also gained attention both from theory and experimental community.  These higher order harmonics can  provide  valuable information about the initial state of the colliding system ~\cite{riseFall, geoFluct1, hydrov3}  and also provide an natural explanation to the ridge phenomena in heavy ion collisions~\cite{derik}.  Measurement of these harmonics in different systems and collisions energy helps to further understand the heavy ion physics in general. 

 In uranium - uranium (U+U) collisions, there is  the potential to produce more extreme conditions of excited matter at higher density and/or greater volume than  is possible using spherical nuclei like gold or lead at the same incident energy~\cite{UUSim}. Uranium has quadrupole deformed shape. So U+U collisions may offer an opportunity to explore wider range of initial eccentricities.  The collisions of special interest are the ``ideal tip-tip" orientation in which the long axes of both deformed nuclei are aligned with the beam axis at zero impact parameter, and the ``ideal body-body" orientation in which the long axes are both perpendicular to the beam axis and parallel to each other at zero impact parameter. The ``ideal tip-tip" and ``ideal body-body" collision events allow to test the prediction of hydrodynamical model(s) by varying the transverse particle density at spatial eccentricity similar to central Au+Au collisions.   
 
 We report here the first  $v_{n}$  measurements in U+U collisions from the STAR  experiment at the Relativistic Heavy Ion Collider (RHIC) at $\sqrt{s_{NN}}=$ 193 GeV. These data were collected during RHIC run-XII in the year 2012. These measurements are presented as a function of transverse momentum ($p_{T}$) and centrality. 
 
 \section{The STAR experiment and Analysis Details}
\begin{figure}[h]
 \center
\includegraphics[width=35pc]{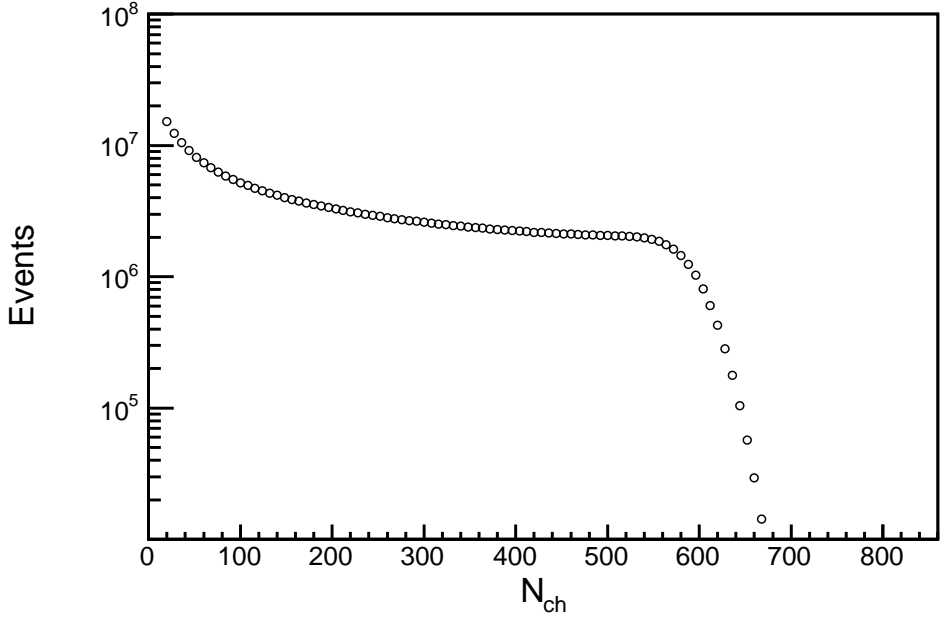}
\caption{ Uncorrected multiplicity distribution with $|\eta| < 0.5 $   in U+U collisions at $\sqrt{s_{NN}}$ = 193 GeV.  Events with $ N_{ch} > 10$ are selected for the present analysis.}
\label{fig1}
\end{figure}

 Data reported in this proceedings were collected in U+U collisions at $\sqrt{s_{NN}}$ = 193 GeV in the year 2012 with a minimum bias trigger  and the ultra central events were taken with a dedicated central trigger.  Main detector used in the present measurements is the Time Projection Chamber (TPC)~\cite{startpc},  the primary tracking device at STAR. TPC has  full azimuthal coverage and uniform acceptance in $\pm 1.0 $ units of pseudorapidity.   The charged particle momenta are measured by reconstructing their trajectories  through the TPC.  The events with the primary collision vertex position along the beam  direction $(V_z)$  within 30 cm of the center of the detector are selected for this analysis.  Event vertex is further required to be in the transverse direction within 2.0 cm from the center of the beam pipe.  Centrality classes in U+U collisions at $\sqrt{s_{NN}}$ = 193 GeV are defined using the number  of charged particle tracks reconstructed in the Time Projection Chamber(TPC) within pseudorapidity $|\eta| < 0.5 $ and passing within 3 cm of interaction  vertex.   The  uncorrected, i.e.  not corrected for acceptance and reconstruction efficiency,  multiplicity ($N_{\rm{ch}}$) distribution for events with a reconstructed primary vertex is shown in Fig. ~\ref{fig1}.  

This analysis  was carried out on tracks that had transverse momenta    $p_{T}  > 0.15$ GeV$/c$, passed within  3 cm of the primary vertex, had at least 15 space points in the main TPC acceptance  $(|\eta| < 1.0)$  and the ratio of the number of actual space points to the maximum possible number of space points for that  track's trajectory was greater than 0.52.

 For any Fourier harmonic \emph{n},  the event flow vector ($Q_n$) and the event plane angle ($\Psi_n$) are defined by~\cite{methodPaper}, 
\begin{equation}  
Q_n \cos (n\Psi_n)  = Q_{nx} = \sum_i w_i\cos (n\phi_i),
\end{equation}
\begin{equation} 
Q_n \sin (n\Psi_n) = Q_{ny} = \sum_i w_i\sin (n\phi_i), 
\end{equation}
\begin{equation} 
\Psi_n  =  \left( \tan^{-1} \frac{Q_{ny}}{Q_{nx}} \right)/n,
\end{equation}
where sums extend over all particles $i$ used in the event plane calculation, and $\phi_i$ and $w_i$ are the laboratory azimuthal angle and the weight for 
the $i$-th particle, respectively.  The event plane vector is  reconstructed from tracks with transverse momentum ($p_{T}$) up to 2 GeV/$c$ and  within $ |\eta| < 1.0$.  For first harmonics(n=1), 
the weight is taken as, 

\begin{equation}
 w_{i} = p_{T} -\frac{ <p_{T}^2>}{<p_{T}>},
  \end{equation}
  and for higher harmonics ($n>1$) the weight is equal to the transverse momentum
  \begin{equation}
  w_{i} = p_{T}.
  \end{equation}

The $<p_{T}>$ and $<p_{T}^{2}>$ represent event averaged quantities.  The choice of this weight corrects the effect of the momentum conservation and also it cancels out the conventional pseudorapidity  asymmetric directed flow~\cite{Luzum, modelPrediction}. 

  We used scalar product method as well as the event plane method to measure the signal. In the scalar product method, the  particles with $ 0.5 <\eta < 1.0  $ were assigned to one subevent and particles with $-1.0 < \eta< -0.5$   to the other subevent,  separated by  $\eta$  gap of 1.0 units between two subevents and at least 0.5 unit with the particle of interest and the subevent to which it is correlated.   In order to remove the acceptance effects we applied recentering correction~\cite{methodPaper} to the flow vectors.  With unit vector defined as $u_{ni}  =  e^{in\phi}$,  the first harmonic coefficient  $v_{n}$ is evaluated as,

\begin{equation}
v_{n}(\eta>0) = \frac{\langle {Q_{na}(\eta<0) u_{ni}^*} \rangle}{\sqrt{\langle Q_{na}Q_{nb} \rangle}}
\label{etasub}
\end{equation}

\begin{equation}
v_{n}(\eta< 0)  = \frac{\langle Q_{nb}(\eta>0) u_{ni}^*) \rangle}{\sqrt{\langle Q_{na}Q_{nb} \rangle}}
\label{etasub}
\end{equation}

In the event plane method, the  particles with $ 0.1 <\eta < 1.0  $ were assigned to one subevent and particles with $-1.0 < \eta< -0.1$   to the other subevent,  so that subevents were separated by  $\eta$   gap of 0.2 units.  Larger pseudorapidity separation was desired  but we are limited by the event plane resolution. In this method, we first evaluate the event plane angle $\Psi_{n}$ from the event plane vector ($Q_{n}$) and  the event plane angle was flattened using a shifting correction method to correct for the detector acceptance effects. The particle correlation  with event plane angle $\Psi_{n}$ is given by, 

\begin{equation}
v_{n}(\eta>0)   =\frac{\langle \cos n(\phi-\Psi_{na}(\eta<0)) \rangle}{\sqrt {\langle \cos n(\Psi_{na}(\eta<0)-\Psi_{nb}(\eta>0)) \rangle}} . 
\label{etasub}
\end{equation}

\begin{equation}
v_{n}(\eta< 0)   = \frac{\langle \cos n(\phi-\Psi_{nb}(\eta>0)) \rangle}{\sqrt {\langle \cos n(\Psi_{na}(\eta<0) -\Psi_{nb}(\eta>0)) \rangle}} . 
\label{etasub2}
\end{equation}

\section{Result and discussion}

Results are presented with only statistical errors. In these studies, contribution  from short range correlation such as Bose-Einstein correlations, coulomb interactions  are studied using a pseudorapidity gap of at least 0.5 units in pseudorapidity between the event vector and the particle of interest using scalar product method. Results from both methods, scalar product with larger psuedorapidity  gap and   event plane method with smaller pseudorapidity gap  are consistent with each other, which suggests that non-flow contribution from short range correlations are small. To reduce the effects from high $p_{T}$ particles in the estimation of the event plane,  we used  $p_{T}$ weight only up to 2 GeV/$c$. The non-flow  contribution from the jets/minijets  is not known and might be a significant  contributor to the systematic uncertainties especially at peripheral collisions. 

\begin{figure*}[htp]
  \centering
  \subfigure[$v_{n}$ integrated in $p_{T} <$ 2 GeV/$c$ and $|\eta|<1.0$ as a function of centrality ]{\includegraphics[scale=0.38]{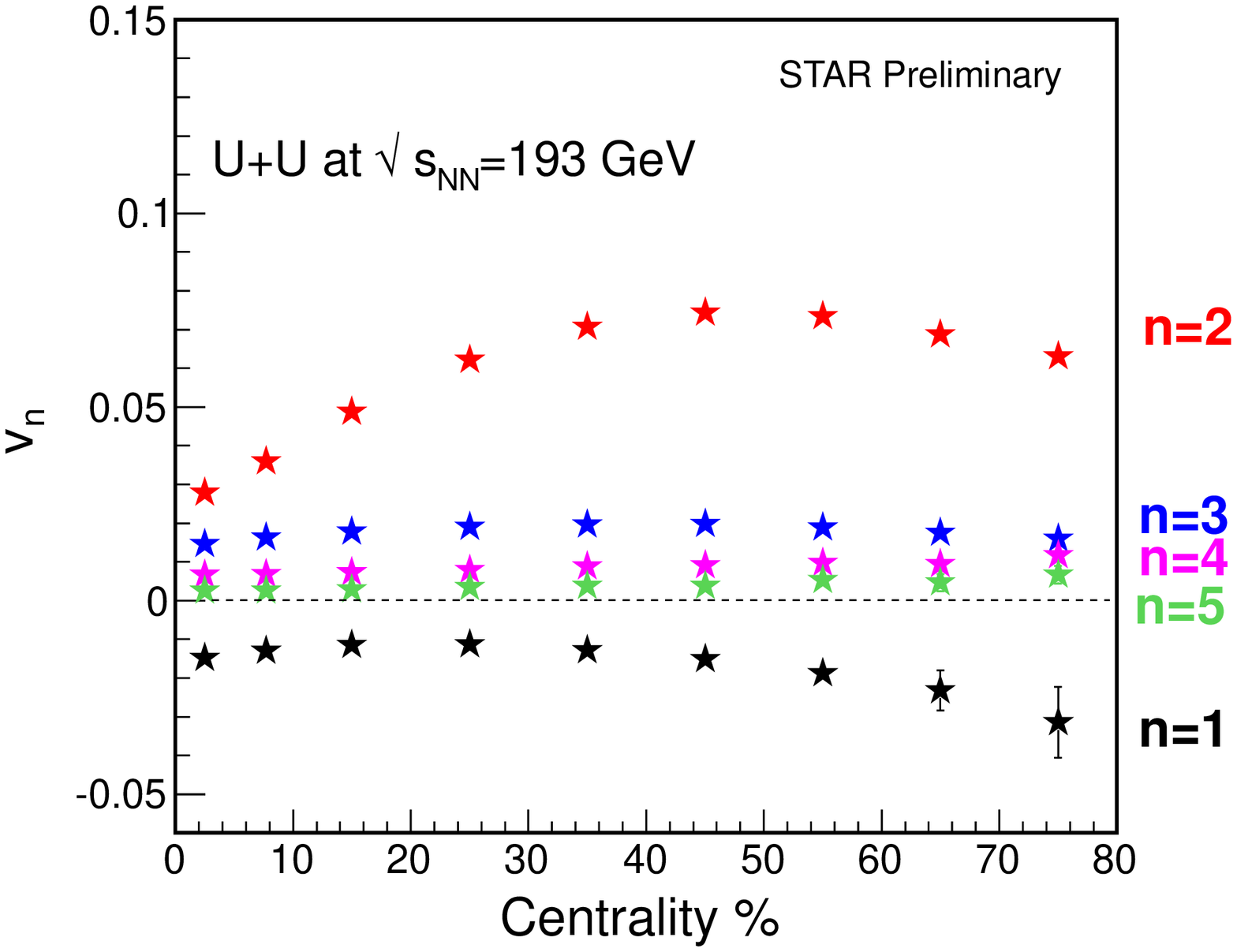}}\quad
  \subfigure[$v_{2}$ as a function of  centrality compared to the prediction based on MC glauber model~\cite{Hiroshi}  ]{\includegraphics[scale=0.38]{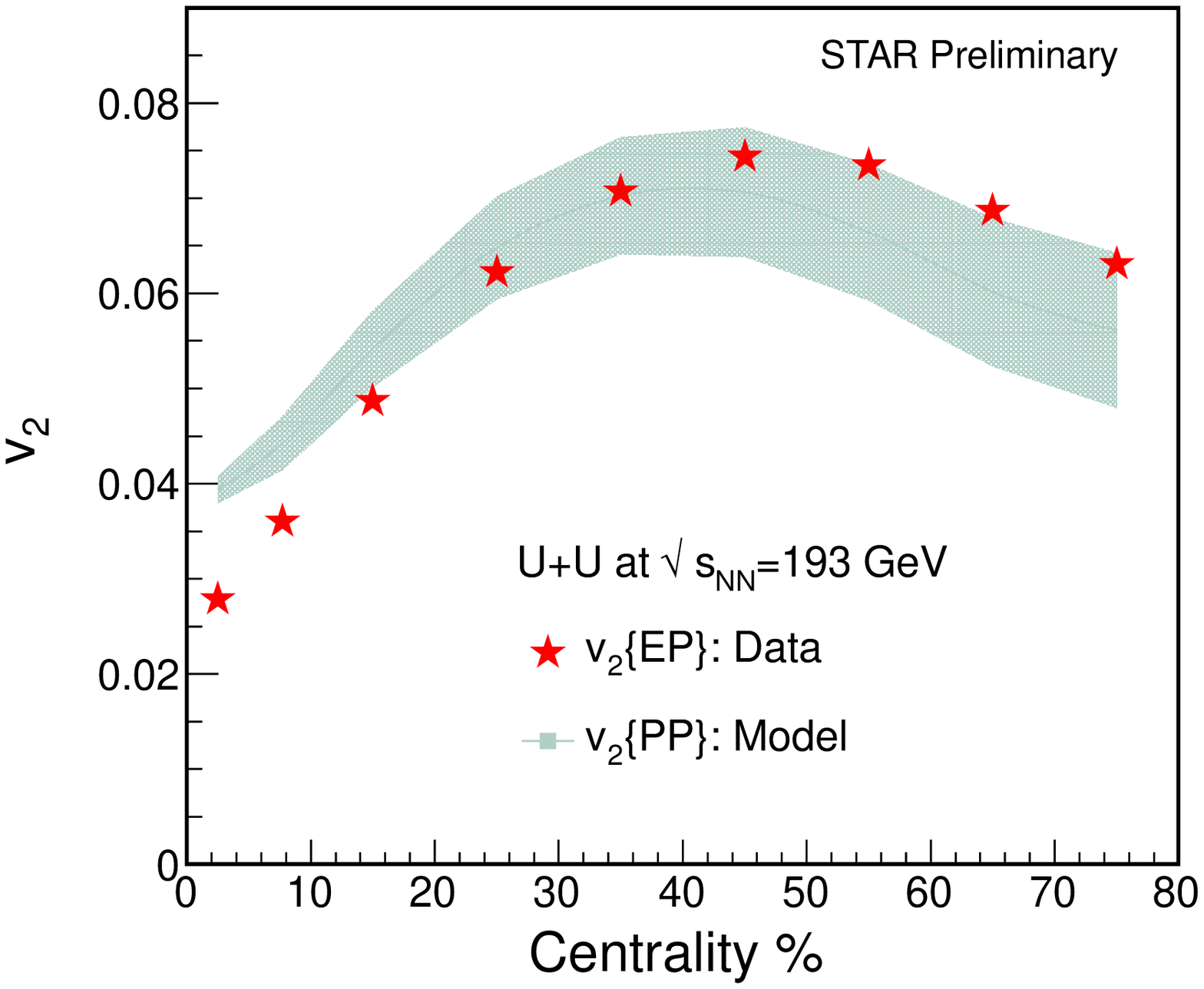}}
  \caption{$v_{n}$ as a function of centrality for U+U Collisions }
    \label{fig2}  
\end{figure*}

          Figure ~\ref{fig2} (a) presents the  $v_{n}$ integrated in transverse momentum $p_{T}$ ($0.15 < p_{T}< 2 GeV/c$) and pseudorapidity $\eta$ ($|\eta|<1.0$) as a function of centrality.  Except for $v_{2}$, we observe very weak centrality dependence.  The initial overlap geometry, which makes the dominant contribution to the elliptic flow, changes from central to peripheral collisions, giving rise to the strong centrality dependence of $v_{2}$. For  $v_{1}$,  centrality dependence is observed only at more peripheral collisions (40-80\%) where non flow contribution  may be significant. The non flow contribution from  the momentum conservation effect ~\cite{modelPrediction} may hot have perfectly suppressed in peripheral collisions in the measurement of $v_{1}$. The centrality dependence of the geometry fluctuations are not yet understood. The weak centrality dependence of the flow coefficients which are believed to originate from geometry fluctuations indicates that geometry fluctuations weakly depend on centrality. 
                   
         Figure ~\ref{fig2} (b) presents the elliptic flow $v_{2}$   as a function of centrality compared with a model prediction based on glauber  based~\cite{globerMC} prediction.  Model curve is for 200 GeV U+ U collisions from reference ~\cite{Hiroshi}. For central collisions, experimental data are lower  than the prediction.  
                          
        The $v_{n}$ of all charged hadrons, for $n =1,\, 2,\, 3,\, 4$ and 5  as a function of $p_{T}$ at various centralities, as shown in Fig. 3.  The data for 0--5\% centrality are shown in the upper panels, and for intermediate centrality (30--40\%) in the lower panels.  All $v_{n}$ measurements show an increasing trend as a function of $p_{T}$.  Except for $v_{2}$, we observe very weak centrality dependence. Also shown are the same measurement  from Au+Au collisions at 200 GeV for the comparisons. We observe the difference in $v_{n}$ for  n=1 and n=2 at 0-5\% central between U+U
and Au+Au collisions. This may hint  to  initial overlap geometry difference in the central collisions between two systems Au+Au and U+U collisions. The difference diminishes in higher harmonics and more peripheral collisions.        
        
 \begin{figure}[h]
\center
\includegraphics[width=35pc]{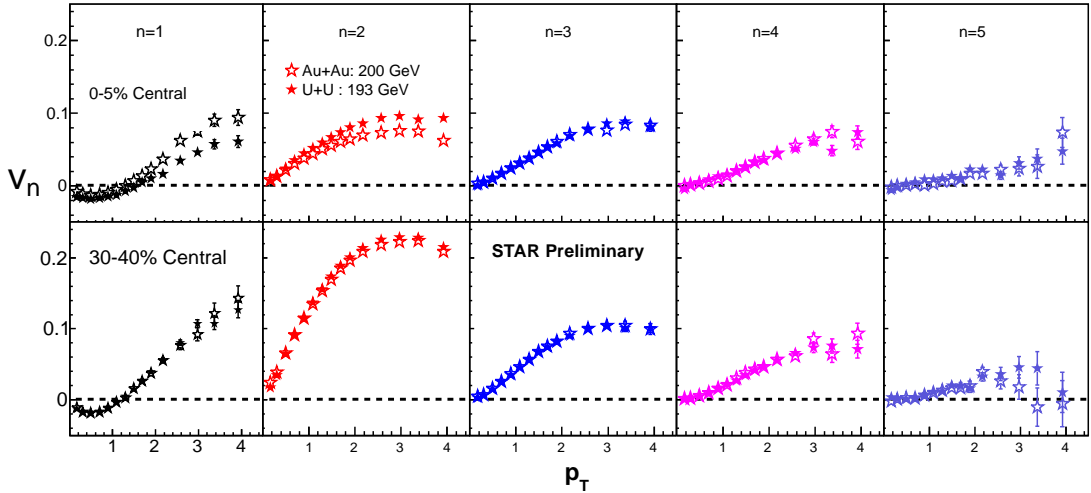}
\caption{$v_{n}$ measurement at 0-5\% central in upper panels and mid central (30-40\%) collisions in the lower panels as a function of $p_{T}$ for U+U collisions at 193 GeV compared with $v_{n}$  measurement for Au+Au collisions at 200 GeV~\cite{QM2012Proceeings}}
\label{fig3}
\end{figure}

\begin{figure*}[htp]
  \centering
  \subfigure[$v_{n}$ as a function of centrality at 0-5\% central collisions] {\includegraphics[scale=0.38]{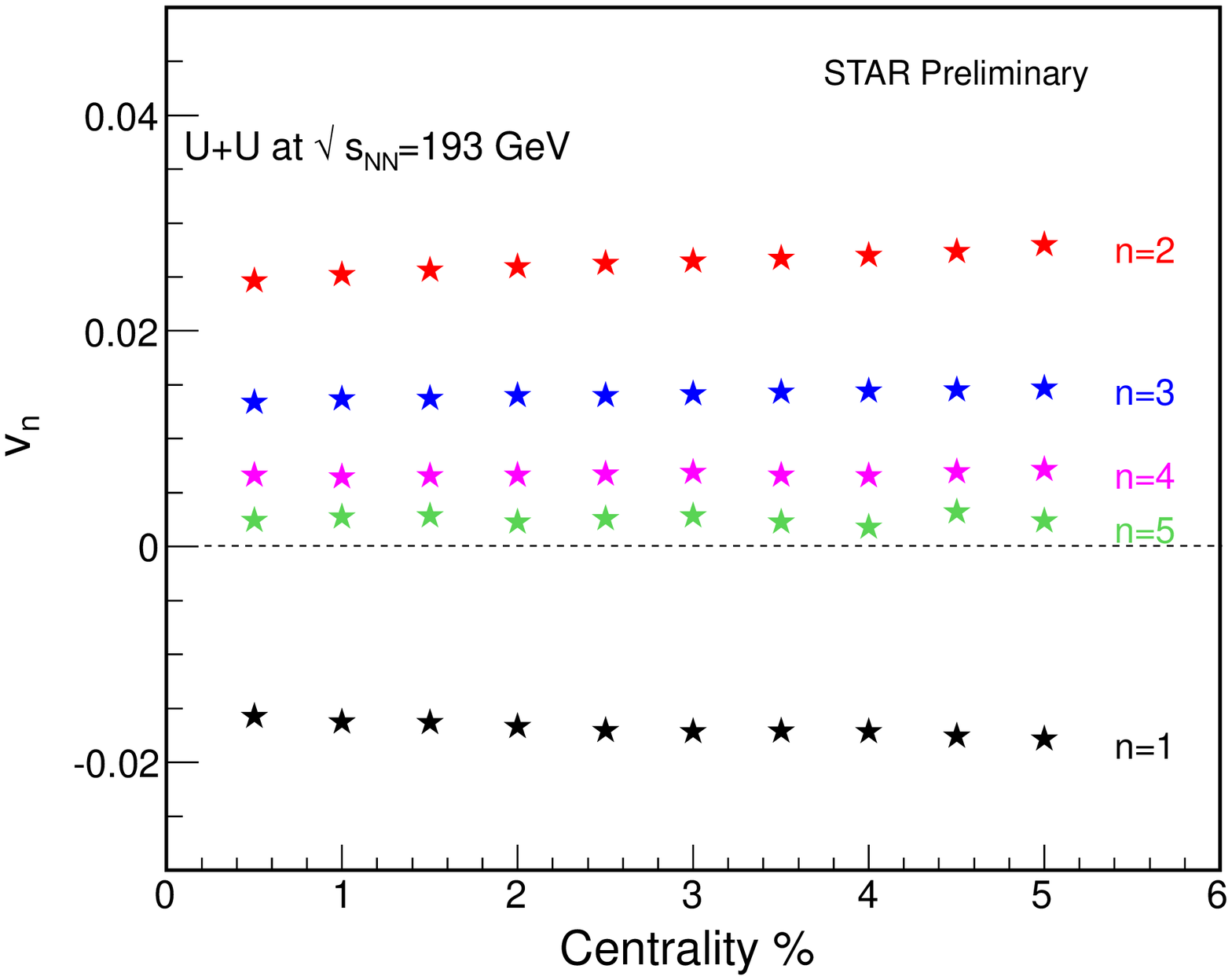}}\quad
  \subfigure[$v_{n}$ as a function of $p_{T}$ at ultra central(0-0.5\%) collisions ]{\includegraphics[scale=0.38]{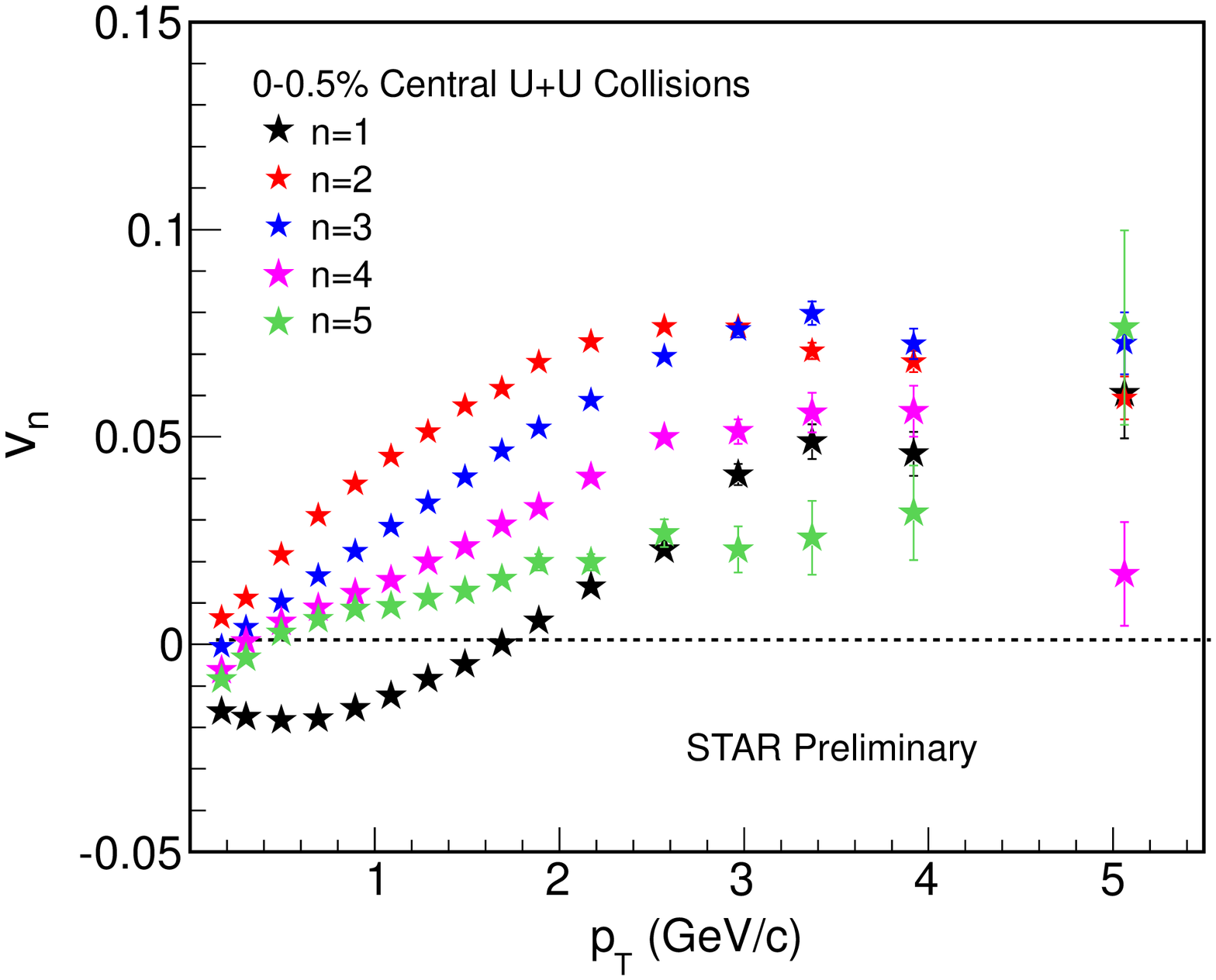}}
  
  \caption{$v_{n}$ as a function of centrality and transverse momentum for ultra central collisions}
  \label{fig4}  
\end{figure*}

We also report measurement of $v_{n}$ for ultra central U+U collisions. The centrality selection is based on reference multiplicity. We subdivide 0-5\% central bin into 10 smaller centrality bins upto most central 0-0.5\% centrality.  The most central  0-0.5\%  collisions should have have dominant contribution from tip-tip collisions.   In Fig.~\ref{fig4} (a), $v_{n}$ as a function of centrality for ultra central collisions is shown. Other than second harmonic coefficient, the $v_{n}$  do not change in this centrality range. We observe small change for $v_{2}$ since it still has some contribution from initial overlap geometry. This observation suggests that there is still a small contribution from body-body collisions. 

  In Fig.~\ref{fig4} (b), the $v_{1}$ signal along with $v_{2}$, $v_{3}$,  $v_{4}$ and  $v_{5}$ a function of transverse momentum up to $p_{T}$  $\sim$ 5 GeV/$c$  is shown for ultra central (0-0.5\%) collisions.  In these most central events we find that higher harmonics are also significant in magnitude compared with  second harmonics in the intermediate transverse momentum. Contributions  from higher harmonics should not be overlooked interpreting  dihadrom correlation data at intermediate transverse momentum. These higher harmonics may offer natural explanation to the novel ridge phenomena in heavy ion collisions~\cite{riseFall}.   Model comparisons with these new data may help us to better understand the medium properties. 
  
\section {Summary}
We report the first measurement of azimuthal anisotropy $v_{n}$ for \textit {n} = 1-5 as a function of transverse momentum $p_{T}$ and centrality in U+U collisions at $\sqrt{s_{NN}}$ = 193 GeV, recorded with the STAR detector at RHIC. Centrality dependence is week for harmonics other than second harmonics.  For higher harmonics and mid central collisions, $v_{n}$(U+U) is similar to $v_{n}$(Au+Au),  the difference appears at central collisions for $v_{1}$ and $v_{2}$. At intermediate $p_{T}$ ranges 3-5GeV/$c$, $v_{n}$ are comparable to the $v_{2}$ signal in ultra central collisions.   Model calculation specially at ultra central collisions  may be useful to constrain the initial condition and  transport coefficient of the medium produced in heavy ion collisions.

\end{document}